\documentclass[aps,prd,reprint,nofootinbib,groupedaddress,preprintnumbers,longbibliography]{revtex4-1}

\usepackage{amsmath,amssymb,mathtools,bm}
\usepackage{graphicx, color}
\usepackage[dvipsnames]{xcolor}
\usepackage{float}
\usepackage[multiple]{footmisc}

\definecolor{c1}{HTML}{802410} 
\definecolor{c2}{HTML}{003262} 
\definecolor{c3}{HTML}{00A598}
\definecolor{c4}{rgb}{1., 0.498039, 0.054902}

\usepackage{tikz}
\usepackage{tkz-euclide}
\usetikzlibrary{decorations.pathmorphing}	
\tikzset{
    v/.style={decorate, decoration={snake, segment length=3mm, amplitude=0.75mm}, draw},
    f/.style={draw,decoration={markings,mark=at position #1 with {\arrow[very thick]{latex}}},postaction={decorate},node contents=#1},
    f/.default=.6,
    fb/.style={draw,decoration={markings,mark=at position #1 with {\arrowreversed[very thick]{latex}}},postaction={decorate},node contents=#1},
    fb/.default=.4,
    fnar/.style={draw},
    g/.style={decorate, draw,  decoration={coil,amplitude=3pt, segment length=3.5pt}},
    s/.style={dashed,draw, postaction={decorate},
        decoration={markings,mark=at position .55 with {\arrow[very thick]{latex}}}},
    sb/.style={dashed,draw, postaction={decorate},
        decoration={markings,mark=at position .55 with {\arrowreversed[draw=black,very thick]{latex}}}},
    snar/.style={dashed,draw,line width =1.25pt},
}
\usetikzlibrary{shapes}	
\tikzset{every picture/.style={line width=1}}
\usepackage{mathrsfs}

\newcommand*\diff{\mathop{}\!\mathrm{d}}

\usepackage{tensor}

\usepackage{orcidlink}

\usepackage{hyperref} 
\hypersetup{
    colorlinks=true,      
    linkcolor=blue,        
    citecolor=blue,        
    filecolor=magenta,     
    urlcolor=blue          
}

\begin{document}

\title{Gravitational Waves from Spectator Scalar Fields}
\author{Marcos A. G. Garcia \orcidlink{0000-0003-3496-3027}}
\email{marcos.garcia@fisica.unam.mx}
\affiliation{Departamento de F\'isica Te\'orica, Instituto de F\'isica, Universidad Nacional Aut\'onoma de M\'exico, Ciudad de M\'exico C.P. 04510, Mexico} 

\author{Sarunas Verner \orcidlink{0000-0003-4870-0826}}
\email{verner.s@ufl.edu}
\affiliation{Institute for Fundamental Theory, Physics Department, University of Florida, Gainesville, FL 32611, USA}
\vspace{0.5cm}

\date{\today}

\begin{abstract}
We propose a novel mechanism for gravitational wave (GW) production sourced by spectator scalar fields during inflation. These fields, while not driving cosmic expansion, generate blue-tilted isocurvature fluctuations that naturally satisfy current CMB constraints at large scales while producing enhanced power spectra at smaller scales accessible to GW detectors. The resulting GW spectrum spans an exceptionally broad frequency range from $10^{-20}$ to $1$ Hz, with amplitudes ranging from $\Omega_{\text{GW}}h^2 \sim 10^{-20}$ to $10^{-12}$ depending on the reheating temperature and spectator field mass. For heavy spectator fields with effective masses near the inflationary Hubble scale $H_I$, the mechanism produces observable signals across multiple detector bands accessible to pulsar timing arrays, space-based interferometers, and ground-based detectors. Our analysis reveals multiple complementary constraints on spectator field parameters. GW-induced limits on the effective number of relativistic species ($\Delta N_{\text{eff}}$) require $m_\chi \gtrsim 0.61 H_I$, stronger than CMB isocurvature bounds alone ($m_\chi \gtrsim 0.54 H_I$). The non-observation of primordial B-modes by \textit{Planck} provides stronger constraints $m_\chi \gtrsim 0.66 H_I$, with projected LiteBIRD sensitivity potentially reaching $m_\chi \gtrsim 0.70 H_I$. This mechanism enables a unique multi-messenger probe of beyond the Standard Model physics during inflation, providing simultaneous constraints on inflationary dynamics, dark matter production, and reheating through current and next-generation GW experiments.
\end{abstract}
\maketitle

\vspace{0.2cm}
\noindent \textbf{Introduction.} 
Cosmic inflation remains the leading theoretical framework addressing several fundamental problems in cosmology, including the horizon problem, the flatness problem, and the origin of large-scale structure~\cite{Guth:1980zm, Starobinsky:1980te, Linde:1981mu, Albrecht:1982wi, Linde:1983gd, Baumann:2007zm}. Despite extensive observational efforts, primordial B-modes, expected imprints of tensor perturbations, have yet to be detected, but they remain a pivotal observational target for testing the inflationary paradigm~\cite{Kamionkowski:1996zd, Seljak:1996gy}. Simple single-field slow-roll models of inflation not only resolve these fundamental issues but also predict an approximately scale-invariant curvature power spectrum, consistent with current observational constraints from the cosmic microwave background (CMB)~\cite{Planck:2018jri} and large-scale structure (LSS) surveys~\cite{Bird:2010mp}. The curvature power spectrum measured by {\em Planck} gives $\Delta^2_\zeta(k_*) \simeq 2.1 \times 10^{-9}$ at the pivot scale $k_* = 0.05 \, \rm{Mpc}^{-1}$~\cite{Planck:2018jri}.

The existence of the Higgs scalar field in the Standard Model (SM)~\cite{Herranen:2014cua, Espinosa:2007qp} strongly motivates the presence of other scalar fields, including spectator fields--fields present during inflation that do not drive the expansion and do not significantly influence its dynamics. Recent studies~\cite{Domenech:2021and, Domenech:2023jve} have shown that large isocurvature fluctuations associated with spectator scalar fields can produce significant and potentially observable GW signals, while~\cite{Ebadi:2023xhq} demonstrated that stochastic effects arising from these fields can independently generate detectable GWs. Related works have explored GW production via parametric resonance~\cite{Cui:2023fbg} and the connection to enhanced curvature perturbations and primordial black hole formation~\cite{Carr:2017edp, Carr:2020xqk}.

Current {\em Planck} constraints limit the isocurvature power spectrum contribution at the CMB pivot scale $k_* = 0.05 \, \rm{Mpc}^{-1}$ to~\cite{Planck:2018jri}:
\begin{equation}
    \label{eq:isocurvaturegen}
    \frac{\Delta^2_{\mathcal{S}}(k_*)}{\Delta^2_{\zeta}(k_*) + \Delta^2_{\mathcal{S}}(k_*)} < 0.038~(\rm{95} \%~C.L.)
\end{equation}
This implies $\Delta^2_{\mathcal{S}}(k_*) < 8.3 \times 10^{-11}$, indicating that isocurvature perturbations must be strongly suppressed at CMB scales. However, it is possible for isocurvature perturbations to have a larger amplitude at smaller scales beyond the reach of current CMB experiments. Current constraints from CMB and LSS primarily limit the curvature power spectrum across scales in the range $k \sim 10^{-4} - 1 \, \rm{Mpc}^{-1}$, while scales smaller than $k > 1 \, \rm{Mpc}^{-1}$ remain largely unconstrained.

In this {\em Letter}, we propose a novel mechanism for GW production sourced by spectator scalar fields, incorporating the complete evolution of isocurvature fluctuations from inflation through reheating to the present day. Our analysis combines analytical approximations with numerical calculations to capture the full cosmological history and reveals that heavy spectator fields naturally produce blue-tilted isocurvature spectra with distinctive observational signatures spanning a broad frequency range from $10^{-20}$ Hz to $1$ Hz. This mechanism provides a powerful probe of beyond the Standard Model physics applicable to various spectator scalar field scenarios, including dark matter candidates~\cite{Chung:1998zb, Chung:1998rq, Peebles:1999fz, Byrnes:2006fr, Markkanen:2018gcw, Kolb:2023ydq, Ling:2021zlj, Garcia:2022vwm, Garcia:2025rut}, curvaton models~\cite{Linde:1996gt, Enqvist:2001zp, Moroi:2001ct, Lyth:2001nq}, axions~\cite{Peccei:1977ur, Peccei:1977hh,Bao:2022hsg, Chen:2023txq}, and supersymmetric theories~\cite{Martin:1997ns, Freedman:2012zz}. As GW astronomy achieves unprecedented multi-band sensitivity, our framework establishes spectator field fluctuations as a novel observational window into inflationary dynamics and the poorly understood reheating epoch, with the broad frequency coverage enabling multi-messenger constraints across pulsar timing arrays (PTAs)~\cite{NANOGrav:2023hvm, EPTA:2015qep, EPTA:2015gke, Weltman:2018zrl}, space-based interferometers~\cite{LISA:2017pwj, Crowder:2005nr, Yagi:2011wg}, and ground-based detectors~\cite{LIGOScientific:2019lzm, Punturo:2010zz, Reitze:2019iox}.

\vspace{0.2cm}
\noindent \textbf{Framework.} 
We adopt natural units, $k_B = \hbar = c = 1$, with metric signature $(-,+,+,+)$. The homogeneous and isotropic Friedmann-Robertson-Walker (FRW) metric is given by $ds^2 = a(\eta)^2(-d\eta^2 + \delta_{ij}dx^idx^j)$, where $a(\eta)$ is the scale factor and $d\eta = dt/a$ denotes conformal time. The general action for our theory is:
\begin{equation}
\begin{aligned}
    \label{eq:action}
    \mathcal{S} = \int \diff^4 x \sqrt{-g}  & \Bigg[ \frac{1}{2}M_P^2 R 
    - \frac{1}{2} (\partial_{\mu} \phi)^2 - V(\phi)  \\
    & - \frac{1}{2} (\partial_{\mu} \chi)^2 - \frac{1}{2} m_{\chi}^2 \chi^2 
    - \frac{1}{2} \sigma \phi^2 \chi^2 \Bigg] \, ,
\end{aligned}    
\end{equation}
where $g = \det g_{\mu \nu}$ is the determinant of the metric, $M_P = 1/\sqrt{8 \pi G_N} \simeq 2.435 \times 10^{18} \, \mathrm{GeV}$ is the reduced Planck mass, and $R$ represents the Ricci scalar. Here, $\phi$ is the inflaton field with potential $V(\phi)$, and $\chi$ is the spectator scalar field, with bare mass $m_\chi$. The interaction term $\sigma \phi^2 \chi^2$ characterizes the coupling between the inflaton and the spectator field. Introducing the canonically normalized field $X \equiv a\chi$, and varying the action with respect to $X$, we obtain the equation of motion:
\begin{equation}
    \label{eq:eom}
   \left( \Box - a^2 m_{\rm eff}^2 \right) X \; = \; 0 \,,~~~~~ m_{\rm eff}^2 = m_{\chi}^2 + \sigma \phi^2 - \frac{1}{6}R\, .
\end{equation}
We expand the field $X$ in Fourier modes as:
\begin{equation}
\label{eq:fourierx}
\hspace{-0.5em} X(\eta, \mathbf{x}) = \int \frac{d^3k}{(2\pi)^{3/2}} e^{-i \mathbf{k} \cdot \mathbf{x}} 
\left[ X_k(\eta) \hat{a}_\mathbf{k} + X_k^*(\eta) \hat{a}_{-\mathbf{k}}^\dagger \right] ,
\end{equation}
where $\mathbf{k}$ is the comoving momentum, with $|\mathbf{k}| = k$, and 
$\hat{a}_\mathbf{k}$ and $\hat{a}_{\mathbf{k}}^\dagger$ are the annihilation and creation operators, respectively, that satisfy the canonical commutation relations: $[\hat{a}_\mathbf{k}, \hat{a}_{\mathbf{k}'}^\dagger] = \delta^{(3)}(\mathbf{k} - \mathbf{k}')$ and $[\hat{a}_\mathbf{k}, \hat{a}_{\mathbf{k}'}] = [\hat{a}_\mathbf{k}^\dagger, \hat{a}_{\mathbf{k}'}^\dagger] = 0$. By substituting the Fourier-decomposed field from Eq.~(\ref{eq:fourierx}) into the equation of motion (\ref{eq:eom}), we obtain the mode equation:
\begin{equation}
\label{eq:modeeq}
X_k'' + \omega_k^2 X_k \; = \; 0 \, , \quad \text{with } \quad \omega_k^2 \; = \; k^2 + a^2 m_{\text{eff}}^2 \, ,
\end{equation}
where primes denote derivatives with respect to conformal time.

Evaluating GW signals from spectator scalar fields requires careful treatment of their gravitational production during and after inflation. 
This process determines the energy density ratio between spectator and inflaton fields, $\rho_\chi/\rho_\phi$, which critically influences isocurvature power spectrum evolution as modes re-enter the horizon.
We solve the mode equations with initial conditions set by the Bunch-Davies vacuum. In the early-time asymptotic limit, $\eta \to -\infty$,
modes deep inside the horizon ($k\gg a H$) evolve in effectively flat Minkowski space with frequency $\omega_k = k$, motivating the vacuum state $\lim_{\eta\to-\infty} X_k(\eta) = \frac{1}{\sqrt{2k}}e^{-ik\eta}$. The comoving energy density of gravitationally produced spectator scalar fields is computed as:~\cite{Ling:2021zlj, Kofman:1997yn, Garcia:2022vwm, Kolb:2023ydq}:
\begin{equation}
  \rho_\chi a^3 = \int_{k_0}^\infty \frac{\diff k}{k} \, \mathcal{E}_k(\eta), ~~\text{with}~~\mathcal{E}_k(\eta) = \frac{k^3}{2 \pi^2} \frac{\omega_k(\eta)}{a(\eta)} n_k\,,
\end{equation}
where the occupation number is given by $n_k = \frac{1}{2\omega_k}|\omega_k X_k - iX'_k|^2$.\footnote{This expression is computed in the late-time asymptotic limit. The complete time-dependent evolution is provided in the Supplemental Material.} We numerically evolve these mode equations to track the complete energy density evolution throughout cosmological history.

For concreteness, we adopt the T-model inflationary potential~\cite{Kallosh:2013maa}:\footnote{The results presented in this {\em Letter} are mostly general and can be applied to any plateau-like inflationary potential.}
\begin{equation}
V(\phi) = \lambda M_P^4 \left[ \sqrt{6} \tanh\left( \frac{\phi}{\sqrt{6} M_P} \right) \right]^2 \, ,
\end{equation}
where the coupling is normalized to {\em{Planck}} observations via $\lambda \simeq 3\pi^2 \Delta^2_{\zeta}(k_*)/N_*^2$. For $N_* = 55$ e-folds, this yields $\lambda \simeq 2 \times 10^{-11}$, spectral tilt $n_s \simeq 0.963$, and tensor-to-scalar ratio $r \simeq 0.004$, which is highly favored by current CMB measurements \cite{Planck:2018jri, Ellis:2021kad}. The parameter $\lambda$ determines the inflaton mass at the potential minimum $V(0)$, with $m_\phi = \sqrt{2\lambda} M_P \simeq 1.6 \times 10^{13} \, \rm{GeV}$. The Hubble parameter at the horizon exit is given by $H_I = 1.5 \times 10^{13} \, \text{GeV} \, (\text{with } \phi_* = 5.35 M_P)$, and at the end of inflation, $H_{\text{end}} = 6.3 \times 10^{12} \, \text{GeV}.$

\vspace{0.2cm}
\noindent \textbf{Curvature and Isocurvature Power Spectra.} 
We analyze metric perturbations in the Newtonian gauge with line element:
\begin{equation}
\label{eq:lineelement}
 ds^2 = a^2(\eta) \left[-(1+2\Phi)d\eta^2 + \left[(1 - 2\Phi)\delta_{ij} +h_{ij} \right]dx^i dx^j \right] \, ,
\end{equation}
where $\Phi$ represents scalar perturbations and $h_{ij}$ denotes tensor perturbations satisfying the transverse-traceless conditions $h^i_{~i} = \partial_i h^{ij} = 0$. Primordial fluctuation dynamics are governed by two gauge-invariant quantities: the curvature perturbation $\zeta = -\Phi - \mathcal{H} \delta\rho/\bar{\rho}'$ on uniform-density hypersurfaces, and the isocurvature perturbation $\mathcal{S} = 3(\zeta_\chi - \zeta_R)$ measuring entropy fluctuations between spectator field and radiation components.\footnote{We assume the inflaton decays into radiation, thus $\zeta_{\phi} = \zeta_R$.} Here $\mathcal{H} = a'/a$ is the conformal Hubble parameter, $\delta\rho$ is the total density perturbation, and $\bar{\rho}'$ is the conformal time derivative of the background energy density. For spectator scalar field $\chi$ with vanishing background value $\langle\chi\rangle = 0$, the effective mass $m_{\text{eff}}^2 = m_\chi^2 + \sigma\phi^2 - \frac{1}{6}R$ determines the amplitude of quantum fluctuations during inflation. When $m_{\rm{eff}} \lesssim H_I$, superhorizon modes are efficiently excited, generating substantial isocurvature perturbations that persist after inflation ends. The statistical properties of these isocurvature fluctuations are captured by the dimensionless power spectrum, $\Delta^2_{\mathcal{S}} \equiv \frac{k^3}{2\pi^2} \mathcal{P}_{\mathcal{S}}(\eta, k)$, where $\mathcal{P}_\mathcal{S}$ is defined through the two-point correlation function $\langle\mathcal{S}(\eta,\mathbf{k})\mathcal{S}(\eta,\mathbf{k}')\rangle = \delta^{(3)}(\mathbf{k+k'})\mathcal{P}_\mathcal{S}(\eta,k)$.

After expressing energy density fluctuations in terms of the canonically normalized field $X = a \chi$, we derive:
\begin{equation}
\begin{aligned}
    &\Delta_{\mathcal S}^2 (\eta, k) = \frac{1}{\bar{\rho}_{\chi}^2(\eta)} \frac{k^3}{a^8 (2\pi)^5 } \int d^3 p \bigg\{|X_p'|^2 |X_q'|^2 \\
    &+ a^4 m_{\rm eff}^4 |X_p|^2 |X_q|^2 + a^2 m_{\rm eff}^2 \big[ (X_p X_p'^*)(X_q X_q'^*) + \text{h.c.}\big] \bigg\} \, ,
\end{aligned}
\end{equation}
where $q = |\mathbf{p-k}|$.\footnote{We note that this represents the late-time asymptotic limit~\cite{Ling:2021zlj, Garcia:2023awt}. The complete evolution of the isocurvature power spectrum requires tracking the full time-dependent expression, which we provide in the Supplemental Material.} The isocurvature power spectrum can be fitted using the analytical form
 \begin{align}
    \label{eq:PSan1}
    \Delta_{\mathcal{S}}^2(k) \; \simeq \; 
   \Lambda^2 \left(\frac{k}{k_{\rm end}}\right)^{2 \Lambda } \ln \left(\frac{k}{H_I} \right) \left(\frac{H_{\rm end}}{H_I}\right)^{2\Lambda}  \,,
\end{align}
where $\Lambda = 2m_{\rm eff}^2/3H_{\rm I}^2$ and $k_{\rm end}=a_{\rm end}H_{\rm end}$ the mode that leaves the horizon at the end of inflation.\footnote{An alternative analytical approximation is given by $\Delta^2_{\mathcal{S}}(k) \simeq \frac{4}{\pi} \Gamma\left(2 - 2\Lambda \right) \sin\left(\pi \Lambda \right) \left( \frac{k}{k_{\text{end}}} \right)^{2 \Lambda}$~\cite{Garcia:2023awt,Pierre:2023jej, Markkanen:2019kpv, Ebadi:2023xhq, Choi:2024bdn, Garcia:2025rut}, though this form does not fully capture the transition from inflation to the matter-dominated epoch.}
\begin{figure*}[t!]
        \centering
        \includegraphics[width=\linewidth]{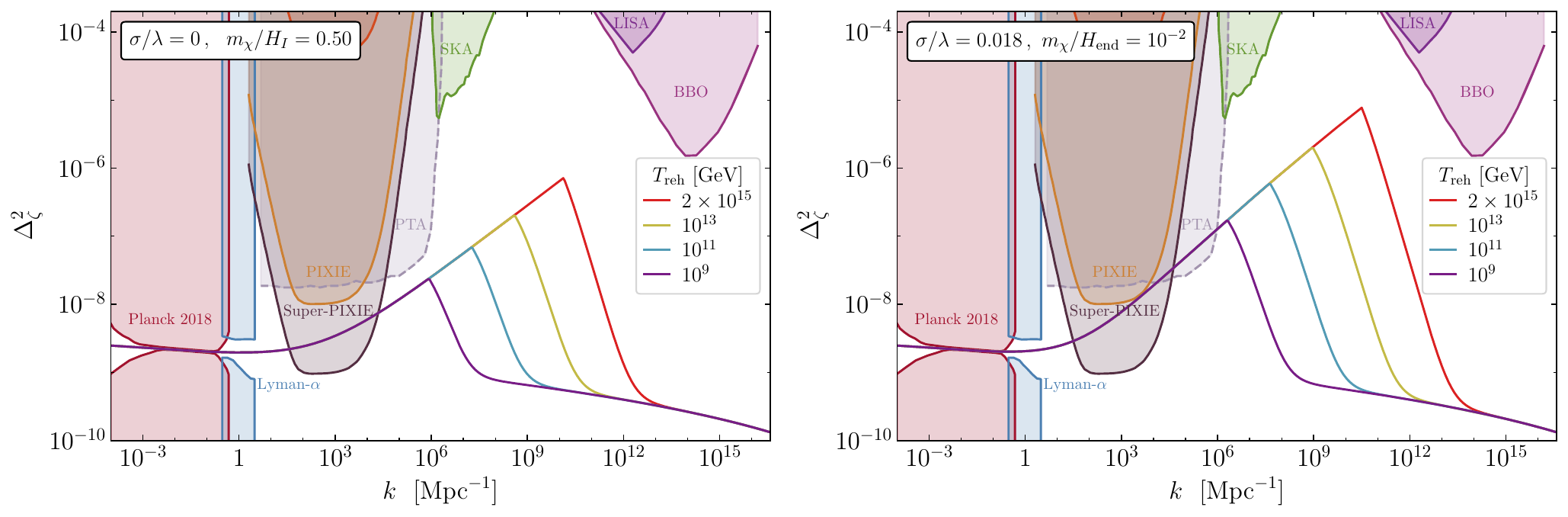}
        \caption{Curvature power spectra for selected spectator field masses and couplings. Each panel shows four reheating temperatures, with the highest corresponding to instantaneous reheating, $a_{\rm reh}=a_{\rm end}$. The spectrum exhibits a characteristic peak at $k_d$ (corresponding to spectator field decay), with red-tilted suppression at smaller scales where $f_{\chi}\ll 1$ and $\Delta^2_{\zeta}\simeq (k_d/k)^2\Delta^2_{\mathcal{S}}/16$. Background curves indicate current and projected sensitivities: CMB spectral distortion experiments (PIXIE, Super-PIXIE~\cite{Chluba:2019kpb, Kogut:2019vqh}), PTA constraints on enhanced dark matter substructure~\cite{Lee:2020wfn}, and GW observatories (SKA~\cite{Carilli:2004nx, Janssen:2014dka, Weltman:2018zrl}, LISA~\cite{LISA:2017pwj, Baker:2019nia}, BBO~\cite{Crowder:2005nr, Corbin:2005ny}).}
        \label{fig:powerspectra}
\end{figure*}

Our analysis reveals a key feature: the isocurvature spectrum exhibits exponential sensitivity to the ratio $m_\text{eff}/H_I$. Small increases in this ratio dramatically suppress isocurvature perturbations at large scales, producing a strongly blue-tilted spectrum that naturally satisfies CMB constraints at observable scales ($k_* = 0.05$ Mpc$^{-1}$) while maintaining significant power at smaller scales where GW production occurs.

After inflation, as the universe transitions to radiation domination, the total curvature perturbation receives contributions from both radiation (originating from inflaton decay) and the spectator field:
 \begin{equation}
    \zeta = \frac{4 \bar{\rho}_R  \zeta_R + 3 \bar{\rho}_\chi \zeta_\chi}{4 \bar{\rho}_R + 3 \bar{\rho}_\chi} 
    \, ,
\end{equation}
where $\zeta_R$ and $\zeta_\chi$ represent the individual curvature perturbations associated with each component. Using the isocurvature perturbation $\mathcal{S} = 3(\zeta_\chi - \zeta_R)$, we can express the total curvature perturbation as:
\begin{equation}
\zeta \; = \; \zeta_R + \frac{\bar{\rho}_\chi}{4 \bar{\rho}_R + 3 \bar{\rho}_\chi} \mathcal{S}\; = \; \zeta_R + \frac{f_\chi}{4 + 3 f_\chi} \, \mathcal{S} \, ,
\end{equation}
where $f_\chi = \rho_\chi/\rho_R$ is the energy density ratio. Since radiation and matter-like spectator fields scale differently with expansion ($\rho_R \propto a^{-4}$, $\rho_\chi \propto a^{-3}$), this ratio increases as $f_\chi \propto a$ until the spectator field decays, creating a scale-dependent modulation of the curvature power spectrum. Assuming statistical independence between inflaton and spectator field fluctuations, the total curvature power spectrum becomes:
\begin{equation}
    \label{eq:deltazetafull}
    \Delta^2_\zeta \; = \; \Delta^2_{\zeta_R} 
    + \left( \frac{\bar{\rho}_\chi}{4 \bar{\rho}_R + 3 \bar{\rho}_\chi} \right)^2 \Delta^2_{\mathcal{S}} \, , 
    \end{equation}
where $\Delta^2_\zeta(k) = \frac{k^3}{2\pi^2} \mathcal{P}_\zeta(k)$ and $\langle \zeta(\mathbf{k}) \zeta(\mathbf{k}') \rangle \equiv \delta^{(3)}(\mathbf{k} + \mathbf{k}') \mathcal{P}_\zeta(k)$, with $\Delta^2_{\zeta_R}(k)$ and $\Delta^2_{\mathcal{S}}(k)$ defined analogously. For modes re-entering the horizon at different epochs, this density ratio creates a scale-dependent enhancement. A mode with wavenumber $k$ re-enters when $k = aH$, allowing us to relate $f_\chi(t)$ to scale $k$ through $f_\chi(t) \; = \; f_\chi(t_d) \left( \frac{k_d}{k} \right)$, where $k_d$ is the wavenumber of the mode re-entering at the spectator field decay time $t_d$. 

This leads to a distinctive piecewise curvature power spectrum~\cite{Ebadi:2023xhq}:
\begin{align}
    \Delta_\zeta^2(k) = 
    \begin{cases}
    	\Delta_{\zeta_R}^2(k) + \left(f_\chi(t_d) \over {4+3 f_\chi(t_d)}\right)^2 \Delta_{\mathcal S}^2(k),~k < k_d,\\
    	 \Delta_{\zeta_R}^2(k) + \left(f_\chi(t_d)(k_d/k) \over {4+3f_\chi(t_d) (k_d/k)}\right)^2 \Delta_{\mathcal S}^2(k),~k > k_d.
    \end{cases}
\end{align}
The resulting spectrum exhibits a characteristic peak at $k_d$, with position depending on reheating temperature, followed by a red-tilted region at smaller scales. These enhanced scalar perturbations directly source GWs through nonlinear interaction, as we demonstrate in the following section.

The complete isocurvature power spectrum for various spectator field masses and couplings $\sigma$ is presented in the Supplemental Material (SM). Figure~\ref{fig:powerspectra} shows the resulting curvature power spectrum $\Delta^2_{\zeta}$ and its dependence on reheating temperature.

\vspace{0.2cm}
\noindent \textbf{Gravitational Waves from Isocurvature Perturbations.} 
Enhanced scalar perturbations from spectator fields source secondary GWs through nonlinear gravitational interactions. The tensor perturbation equation in Fourier space is:
\begin{equation}
    \label{eq:hijgensol}
    h_{\lambda}''(\eta, \mathbf{k}) + 2\mathcal{H}  h_{\lambda}'(\eta,\mathbf{k}) + k^2 h_{\lambda}(\eta, \mathbf{k})  \; = \; S_{\lambda}(\eta, \mathbf{k}) \, ,
\end{equation}
where $h_\lambda$ represents the two polarization modes of GWs, and
\begin{equation}
h_{ij}(\eta, \mathbf{x}) \; = \; \sum_\lambda \int \frac{d^3k}{(2\pi)^{3/2}} e^{-i \mathbf{k} \cdot \mathbf{x}} \varepsilon_{ij}^\lambda(\mathbf{k}) h_\lambda(\eta, \mathbf{k}) \, ,
\end{equation}
where $\varepsilon_{ij}^\lambda(\mathbf{k})$ is the corresponding polarization tensors. The source term directly couples to the isocurvature perturbations through the gauge-invariant Poisson equation $\nabla^2\Phi = \delta\rho_{\chi}/2M_P^2$, with $S_{\lambda}(\eta, \mathbf{k}) \sim \int d^3p \, p^2 |\mathbf{k} - \mathbf{p}|^2  \, \Phi(\eta, \mathbf{p}) \Phi(\eta, \mathbf{k} - \mathbf{p}) \sim \int d^3p \, \delta\rho_{\chi}(\eta, \mathbf{p})  \delta\rho_{\chi}(\eta, \mathbf{k} - \mathbf{p})$. The gravitational wave power spectrum is defined as: $\langle h_\lambda(\eta, \mathbf{k}) h_{\lambda'}(\eta, \mathbf{k}') \rangle = \delta_{\lambda \lambda'} \delta^{(3)}(\mathbf{k} + \mathbf{k}') \mathcal{P}_\lambda(\eta, k)$ with dimensionless spectrum $\Delta^2_\lambda(\eta, k) = k^3\mathcal{P}_\lambda(\eta, k)/2\pi^2$. 
The present-day gravitational wave energy density is: 
\begin{equation}
   \Omega_{\rm{GW}} h^2 \; = \; \frac{1}{12} \left( \frac{k}{a_0 H_0} \right)^2 \Delta_h^2  \, ,
\end{equation}
where $a_0$ is the present-day scale factor, $H_0 = 100 \, h \, \text{km/s/Mpc}$ is the present-day Hubble parameter, and $\Delta_h^2 = \sum_\lambda \Delta_\lambda^2$ represents the total power across both polarization states. 

One can then show that if one imposes statistical homogeneity and isotropy on the perturbation $\delta \rho_{\chi}$ that arises in the source term $S_{\lambda}$, it leads to $\Delta_h^2 \sim \langle h_\lambda h_\lambda \rangle \sim \langle S_{ \lambda} S_{\lambda} \rangle \sim \langle \delta \rho_{\chi} \delta \rho_{\chi} \delta \rho_{\chi} \delta \rho_{\chi} \rangle \sim \langle \delta \rho_{\chi} \delta \rho_{\chi} \rangle^2$, where we assume that there is no significant presence of primordial non-Gaussianities as perturbative corrections to the Gaussian statistics~\cite{Garcia-Saenz:2022tzu}. Therefore, we derive:
\begin{equation}
\begin{aligned}
    &\Delta_h^2 \sim  k^2\int_0^{\infty} dp\, p\int_{|k-p|}^{k+p} \\
    &\times dq\,  q\,\frac{(k^4-2k^2(p^2+q^2)+(p^2-q^2)^2)^2}{p^7q^7}\,\Delta_{\mathcal{S}}^2(p)\Delta_{\mathcal{S}}^2(q) \, .
\end{aligned}    
\end{equation}
This expression directly relates the GW spectrum to the isocurvature power spectrum, creating a distinctive observational signature of spectator field dynamics. The complete derivation is provided in the SM.

\vspace{0.2cm}
\noindent \textbf{Results and Discussion.} 
We demonstrate that heavy spectator fields with effective masses near the Hubble scale ($m_{\rm eff} \sim H_I$) naturally generate blue-tilted isocurvature spectra with spectral indices $\Lambda \simeq 2 m^2_{\rm eff}/3H_I^2$ (typically $\Lambda \simeq 0.25 - 0.50$), producing observable GWs across frequencies from $10^{-20}$ to $1$ Hz. Our analysis combines analytical approximations with numerical evolution of spectator field dynamics, employing normal-ordered energy density operators and regularized momentum integrals to capture the complete gravitational production mechanism.

Two distinct phenomenological regimes emerge depending on reheating temperature $T_{\rm reh}$ and spectator field evolution. In curvaton-like scenarios with high reheating temperatures ($T_{\text{reh}} \geq 10^7\,\text{GeV}$), spectator fields decay before matter-radiation equality, leading to:
\begin{equation}
\Omega_{\text{GW}} h^2 \propto 
\begin{cases}
k^{4\Lambda - 2} \ln^2(k/H_I), & k \ll a_d H_d \, ,\\
k^{4\Lambda - 4} \ln^2(k/H_I), & k \gg a_d H_d \, ,
\end{cases}
\end{equation}
with GW amplitude scaling $\propto T_{\text{reh}}^{4/3}$. For dark matter production scenarios ($T_{\text{reh}} \lesssim 10\,\text{TeV}$), stable spectator fields produce suppressed amplitudes with scaling $\propto T_{\text{reh}}^{-2/3}$ and:
\begin{equation}
\Omega_{\text{GW}} h^2 \propto 
\begin{cases}
k^{4\Lambda - 2} \ln^2(k/H_I), & k \ll a_{\text{reh}} H_{\text{reh}} \, , \\
k^{4\Lambda - 6} \ln^2(k/H_I), & k \gg a_{\text{reh}} H_{\text{reh}} \, .
\end{cases}
\end{equation}
Both scenarios share identical large-scale behavior but differ at small scales, where gravitational production shows stronger suppression due to the absence of spectator field decay enhancement.

\begin{figure}[t!]
\centering
    \includegraphics[width=0.45 \textwidth]{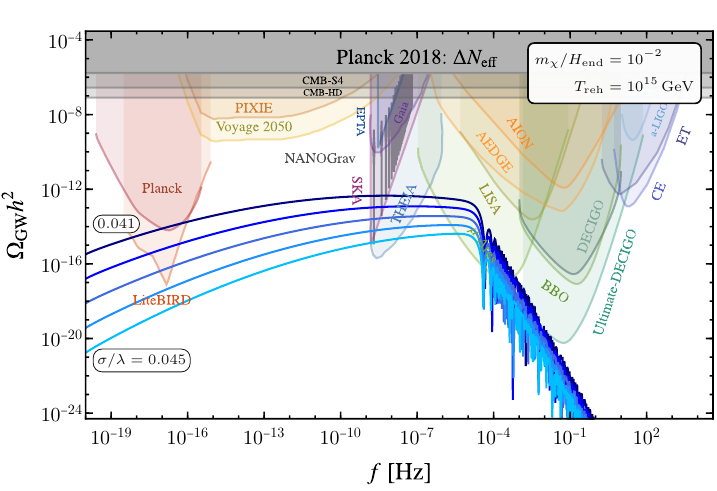}
    \includegraphics[width=0.45 \textwidth]{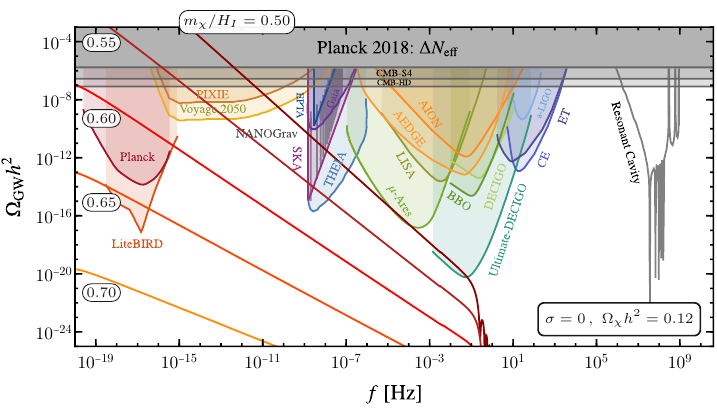}
    \caption{Gravitational wave spectra for two spectator field scenarios. \textit{Top}: Unstable spectator field decaying near matter-radiation equality in a high reheating scenario ($T_{\text{reh}} \simeq 10^{15}\,\text{GeV}$). \textit{Bottom}: Stable spectator field constituting dark matter through purely gravitational production. Background curves show current and projected sensitivities for CMB experiments~\cite{Planck:2018vyg, Abazajian:2019eic, CMB-HD:2022bsz, Hazumi:2019lys, Kogut:2019vqh, Alvarez:2019rhd}, PTAs~\cite{NANOGrav:2023hvm, EPTA:2015qep, EPTA:2015gke, Weltman:2018zrl},  and space/ground-based interferometers~\cite{LIGOScientific:2019lzm, Punturo:2010zz, Reitze:2019iox, Badurina:2021rgt, AEDGE:2019nxb, Garcia-Bellido:2021zgu, LISA:2017pwj, Yagi:2011wg, Crowder:2005nr, Sesana:2019vho}.}
    \label{fig:GW_inst}
\end{figure}

The GW spectra exhibit rich phenomenology depending on the inflaton-spectator coupling. For purely gravitational production ($\sigma = 0$), spectra are universally red-tilted for potentially detectable amplitudes. Non-zero coupling introduces additional complexity: small couplings ($\sigma/\lambda \lesssim 0.03$) maintain red-tilted spectra, intermediate values ($0.03 \lesssim \sigma/\lambda \lesssim 0.05$) produce peaked spectra with maxima around $f \simeq 10^{-6}\,\text{Hz}$, while larger couplings ($\sigma/\lambda \geq 0.05$) generate blue-tilted infrared behavior for $m_{\chi}=10^{-2}H_{\rm end}$). The top panel of Fig.~\ref{fig:GW_inst} shows that for instantaneous reheating at $T_{\text{reh}} = 10^{15}\,\text{GeV}$ and $m_\chi/H_{\text{end}} = 10^{-2}$, the coupling range $0.041 \lesssim \sigma/\lambda \lesssim 0.044$ produces correlated signatures across CMB experiments, PTAs, and space-based interferometers.

\begin{figure}[t!]
\centering
    \includegraphics[width=0.5 \textwidth]{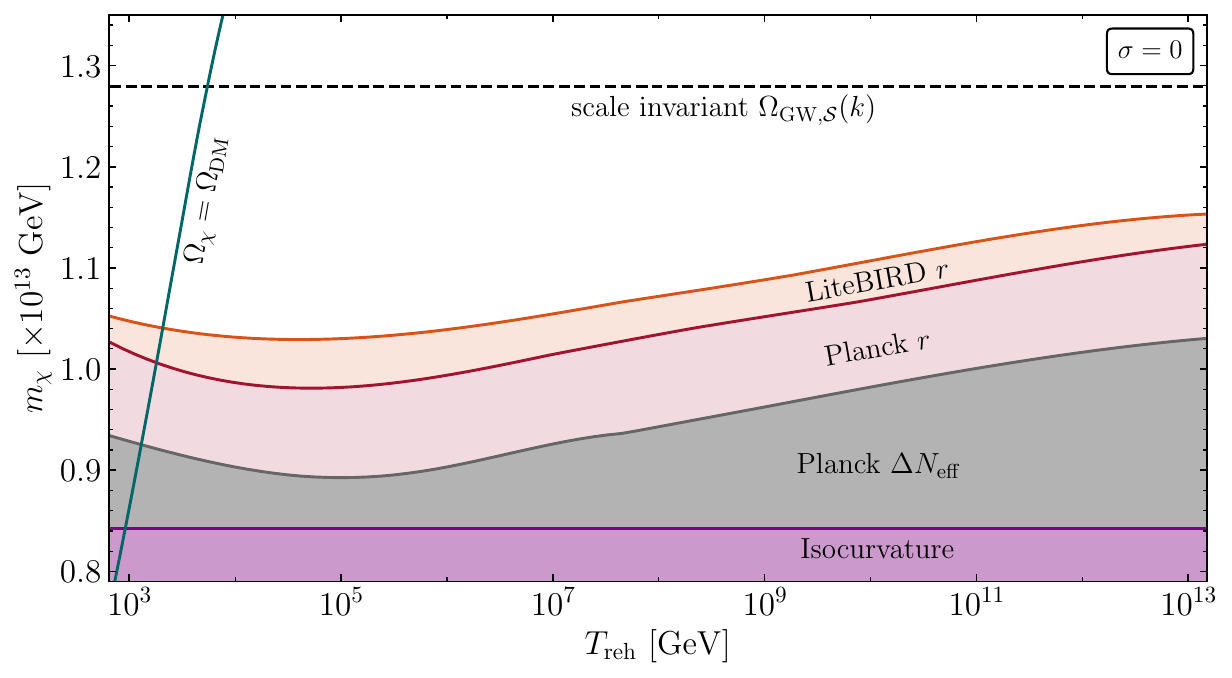}
    \caption{Parameter space constraints for purely gravitational production of heavy spectator fields with $f_{\chi}(t_{\rm d})=1$. Excluded regions: CMB isocurvature bounds (purple, $m_{\chi} \lesssim 0.54 H_I \simeq 8.4 \times 10^{12} \, \rm{GeV}$), $\Delta N_{\rm eff}$ constraints from GW backgrounds (gray), and non-observation of primordial B-modes by {\textit{Planck}}/BICEP2/Keck (red). Orange region shows projected LiteBIRD sensitivity. The teal curve marks the dark matter relic abundance boundary ($\Omega_{\chi}h^2 = 0.12$), with overproduction (underproduction) to the right (left).}
    \label{fig:GW_parspace}
\end{figure}

For dark matter production scenarios, the bottom panel of Fig.~\ref{fig:GW_inst} demonstrates that purely gravitational production can account for the observed relic abundance $\Omega_{\text{DM}} h^2 = 0.12$ while generating detectable GW signals. Spectator masses $m_\chi/H_I = \{0.50, 0.55, 0.60, 0.65, 0.70\}$ correspond to reheating temperatures $T_{\text{reh}} = \{700, 982, 1360, 1867, 2546\}\,\text{GeV}$, respectively. This dark matter boundary (teal curve in Fig.~\ref{fig:GW_parspace}) separates parameter regions producing over- or under-abundance relative to observations.

Parameter space constraints from multiple observations are summarized in Fig.~\ref{fig:GW_parspace}. CMB isocurvature bounds exclude spectator masses $m_\chi \lesssim 0.54 H_I \simeq 8.4 \times 10^{12}\,\text{GeV}$, while our GW analysis reveals stronger constraints through $\Delta N_{\text{eff}} \lesssim 0.28$ bounds that exclude combinations of small masses and high reheating temperatures~\cite{Planck:2018vyg}, requiring $m_{\chi} \gtrsim 0.61 H_I \simeq 9.2 \times 10^{13}\,\text{GeV}$ for $T_{\rm reh} \simeq 10^{8} \, \rm{GeV}$. The non-observation of primordial B-modes by {\textit{Planck}}/BICEP2/Keck~\cite{Planck:2018vyg, BICEP:2021xfz} further constrains $m_\chi \gtrsim 0.66 H_I \simeq 9.9 \times 10^{12}\,\text{GeV}$ except at very low reheating temperatures. Future CMB observatories like LiteBIRD~\cite{Hazumi:2019lys} will probe remaining viable parameter space, with projected sensitivities reaching $m_{\chi} \gtrsim 0.7H_I \simeq 1.1 \times 10^{13}\,\text{GeV}$.

Our analysis reveals that spectator field isocurvature perturbations provide a novel probe of the elusive reheating epoch, as GW amplitudes depend strongly on reheating temperature. Future detection or non-detection could yield the first direct observational constraints on post-inflationary thermalization. The broad frequency coverage, spanning 1 nHz (accessible to SKA~\cite{Weltman:2018zrl}, NANOGrav~\cite{NANOGrav:2023hvm}, and EPTA~\cite{EPTA:2015qep, EPTA:2015gke}) through 1–10 Hz (e.g., $\mu$-Ares~\cite{Sesana:2019vho}, AION~\cite{Badurina:2021rgt}, AEDGE~\cite{AEDGE:2019nxb}) to 1 kHz (Einstein Telescope~\cite{Punturo:2010zz}, Cosmic Explorer~\cite{Reitze:2019iox}), makes multi-band GW astronomy essential for characterizing these signatures.

Spectator fields bridge multiple frontiers in fundamental physics, connecting inflation with dark matter production while probing non-standard thermal histories and potential beyond-Standard-Model coupling structures. We identify parameter space ``sweet spots" (e.g., $m_\chi/H_I \simeq 0.6$-$0.7$ for purely gravitational production with $\sigma = 0$) where signals satisfy current constraints yet remain detectable by future experiments, providing concrete observational targets.

As next-generation experiments achieve unprecedented sensitivity, from CMB-HD targeting $\Delta N_{\text{eff}} = 0.014$ and LiteBIRD probing $r > 0.001$~\cite{Hazumi:2019lys} to BBO reaching $\Omega_{\text{GW}} h^2 \simeq 10^{-16} - 10^{-17}$~\cite{Crowder:2005nr}, these signals could provide unique insights into beyond the Standard Model physics at $10^{13}-10^{15}\,\text{GeV}$ energy scales, dark matter production mechanisms, and the thermal history between inflation and Big Bang Nucleosynthesis. The blue-tilted spectra naturally avoids problematic infrared divergences that plague inflation scenarios, while our combined analytical-numerical methodology establishes a robust framework for interpreting data from this emerging multi-messenger window into the primordial universe.

\vspace{0.2cm}
\noindent \textbf{Acknowledgments.} 
The authors would like to thank Andrew Long for a detailed discussion on the isocurvature power spectrum, and also thank Jeff Dror, David Kaiser, Rocky Kolb, Soubhik Kumar, Evan McDonough, and Vincent Vennin for helpful conversations. The work of S.V. was supported in part by DOE grant grant DE-SC0022148 at the University of Florida. MG was supported by the DGAPA-PAPIIT grants IA103123 and IA100525 at UNAM, and the SECIHTI (formerly CONAHCYT) “Ciencia de Frontera” grant CF-2023-I-17.

\bibliography{references}


\clearpage

\onecolumngrid

\newpage

\widetext
 \begin{center}
   \textbf{\large SUPPLEMENTAL MATERIAL \\[.2cm] ``Gravitational Waves from Spectator Scalar Fields''}\\[.2cm]
  \vspace{0.05in}
  {Marcos A. G. Garcia and Sarunas Verner}
\end{center}
\setcounter{equation}{0}
\setcounter{figure}{0}
\setcounter{table}{0}
\setcounter{page}{1}
\setcounter{section}{0}
\makeatletter
\renewcommand{\thesection}{S-\Roman{section}}
\renewcommand{\theequation}{S-\arabic{equation}}
\renewcommand{\thefigure}{S-\arabic{figure}}



\section{Regularization and Renormalization of Energy Density}
\label{app:A}
This section details the regularization and renormalization procedure for the spectator field energy density $\rho_\chi(\eta)$, addressing ultraviolet (UV) divergences that arise from high-momentum field modes in the curved spacetime background. The treatment requires both systematic regularization of divergent integrals and renormalization through normal ordering with respect to the appropriate vacuum state.

We begin with the stress-energy tensor for the spectator field $\chi$, derived from the action via $\sqrt{-g} T_{\mu\nu}^{(\chi)} \; = \; 2\delta(\sqrt{-g} \mathcal{L}_\chi) / \delta g^{\mu\nu}$~\cite{Birrell:1982ix}:
\begin{equation}
T_{\mu\nu}^{(\chi)} \; = \; \nabla_\mu \chi \nabla_\nu \chi - \frac{1}{2} g_{\mu\nu} g^{\rho\sigma} (\nabla_\rho \chi)(\nabla_\sigma \chi) + \frac{1}{2} g_{\mu\nu} \left( m_\chi^2 + \sigma \phi^2 \right) \chi^2 \,.
\end{equation}
In the FRW background, the $00$ component becomes:
\begin{equation}
T_{00}^{(\chi)}(\eta, \mathbf{x}) \; = \; \frac{1}{2a^2} \left[\chi'^2 + (\nabla \chi)^2 + a^2 \left(m_{\chi}^2 + \sigma \phi^2 \right) \chi^2 \right].
\end{equation}
Using the field redefinition $X = a \chi$ and the relation $\chi' = a^{-1}(X' - \mathcal{H} X)$, we obtain:
\begin{equation}
\label{eq:T00gen}
T_{00}^{(\chi)}(\eta, \mathbf{x}) \; = \; \frac{1}{2a^4} \left[ (X')^2 + (\nabla X)^2 - 2 \mathcal{H} X X' + \left(\mathcal{H}^2 - \frac{a^2 R}{6} \right) X^2 + a^2 m_{\text{eff}}^2 X^2 \right] \,,
\end{equation}
where $\mathcal{H} = a'/a$ is the conformal Hubble parameter and $R$ is the Ricci scalar.

We expand the rescaled field $X$ in Fourier modes:
\begin{equation}
\label{eq:xfourier}
X(\eta, \mathbf{x}) \; = \; \int \frac{d^3 k}{(2\pi)^{3/2}} e^{-i \mathbf{k} \cdot \mathbf{x}} \left[ X_k(\eta) \hat{a}_k + X_k^*(\eta) \hat{a}_{-k}^\dagger \right] \, ,
\end{equation}
where $\hat{a}_k$ and $\hat{a}_k^\dagger$ satisfy canonical commutation relations:
\begin{equation}
[\hat{a}_k, \hat{a}_{k'}^\dagger] = \delta^{(3)}(\mathbf{k} - \mathbf{k'})\, ,~\text{and}~[\hat{a}_k, \hat{a}_{k'}] = [\hat{a}_k^\dagger, \hat{a}_{k'}^\dagger] = 0 \, .
\end{equation}
The mode functions $X_k(\eta)$ evolve according to:
\begin{equation}
\label{eq:eomX}
X_k''(\eta) + \omega_k^2(\eta) X_k(\eta) \; = \; 0 \, ,
\end{equation}
with $\omega_k^2(\eta) = k^2 + a^2(\eta) m_{\text{eff}}^2(\eta)$ and Wronskian normalization $X_k X_k^{\prime *} - X_k^* X_k' = i$.

To handle ultraviolet divergences, we employ adiabatic regularization based on the WKB expansion. We introduce the adiabatic basis functions:
\begin{equation}
f_k(\eta) \; \equiv \; \frac{e^{-i \int^\eta \omega_k d\eta'}}{\sqrt{2 \omega_k}} \,,
\end{equation}
and express the mode functions through Bogoliubov coefficients:
\begin{equation}
X_k(\eta) = \alpha_k(\eta) f_k(\eta) + \beta_k(\eta) f_k^*(\eta) \,,
\end{equation}
with the corresponding derivative
\begin{equation}
    X_k'(\eta) \; = \;  -i \omega_k \left[\alpha_k(\eta) f_k(\eta) - \beta_k(\eta) f_k^*(\eta)\right] \, ,
\end{equation}
where $|\alpha_k|^2 - |\beta_k|^2 = 1$ preserves canonical commutation relations. We define time-dependent ladder operators through the Bogoliubov transformation:
\begin{equation}
\begin{aligned}
\label{eq:rotbog1}
A_k(\eta) &= \alpha_k(\eta) \hat{a}_k + \beta_k^*(\eta) \hat{a}_{-k}^\dagger \, , \\
A_{-k}^\dagger(\eta) &= \beta_k(\eta) \hat{a}_k + \alpha_k^*(\eta) \hat{a}_{-k}^\dagger \,.
\end{aligned}
\end{equation}
The adiabatic vacuum $|0\rangle$ is defined by $A_k |0\rangle = 0$ for all $\mathbf{k}$. The vacuum is normalized such that $\langle 0|0 \rangle = 1$.

The inverse Bogoliubov transformation allows us to express the original operators in terms of the adiabatic ladder operators:
\begin{equation}
\begin{aligned}
\hat{a}_{\mathbf{k}} &= \alpha_k^* A_k - \beta_k^* A_{-k}^\dagger \, , \\
\hat{a}_{-\mathbf{k}} &= \alpha_k^* A_{-k} - \beta_k^* A_k^\dagger \,.
\end{aligned}
\end{equation}
Substituting these relations into the field expansion, we can express the rescaled field and its derivative directly in terms of the adiabatic basis:
\begin{equation}
\begin{aligned}
X(\eta, \mathbf{x}) &= \int \frac{d^3 k}{(2\pi)^{3/2}} e^{-i \mathbf{k} \cdot \mathbf{x}} \left[f_k(\eta) A_k + f_k^*(\eta) A_{-k}^{\dagger} \right] \,, \\
X'(\eta, \mathbf{x}) &= -i \int \frac{d^3 k}{(2\pi)^{3/2}} e^{-i \mathbf{k} \cdot \mathbf{x}} \omega_k \left[f_k(\eta) A_k - f_k^*(\eta) A_{-k}^{\dagger} \right]\,.
\end{aligned}
\end{equation}

Using the Bogoliubov transformation with rotated ladder operators $A_k$, $A_{-k}^\dagger$, $A_k'$, and $A_{-k'}^\dagger$, the $00$ component of the stress-energy tensor for the spectator field can be expressed as:
\begin{equation}
    T_{00}^{(\chi)}(\eta, \mathbf{x}) \; = \; \frac{1}{2a^4} \int \frac{d^3 k d^3k'}{(2\pi)^3} \left[ \mathcal{B}_1(k, k') A_k A_{k'} + \mathcal{B}_2(k, k') A_k A_{-k'}^{\dagger} + \mathcal{B}_3(k, k') A_{-k}^{\dagger} A_{k'} + \mathcal{B}_4(k, k') A_{-k}^{\dagger} A_{-k'}^{\dagger} \right] \, .
\end{equation}
The coefficient functions $\mathcal{B}_i(k, k')$ encode the field dynamics and curvature coupling effects:
\begin{align}
    &\mathcal{B}_1(k,k') \; = \; \left[ \left( i \, \omega_k + \mathcal{H} \right) \left( i \, \omega_{k'} +  \mathcal{H} \right) - k k' + a^2 m_{\text{eff}}^2 - \frac{a^2 R}{6}  \right] f_k(\eta) f_{k'}(\eta) \, , \\
    &\mathcal{B}_2(k,k') \; = \; \left[ \left( -i \, \omega_k -  \mathcal{H} \right) \left( i \, \omega_{k'} -  \mathcal{H} \right) - k k' + a^2 m_{\text{eff}}^2 - \frac{a^2 R}{6} \right] f_k (\eta)f_{k'}^*(\eta)\, , \\
    &\mathcal{B}_3(k,k') \; = \;  \left[ \left( i \, \omega_k -  \mathcal{H} \right) \left( -i \, \omega_{k'} - \mathcal{H} \right) - k k' + a^2 m_{\text{eff}}^2  - \frac{a^2 R}{6} \right] f_k^*(\eta) f_{k'}(\eta)
    \, , \\
    &\mathcal{B}_4(k,k') \; = \; \left[ \left( i \, \omega_k - \mathcal{H} \right) \left( i \, \omega_{k'} -\mathcal{H} \right) - k k' + a^2 m_{\text{eff}}^2  - \frac{a^2 R}{6}  \right] f_{k'}^*(\eta) f_k^*(\eta)
    \, .
\end{align}
This expression contains ultraviolet divergences that arise from high-momentum mode contributions where the adiabatic approximation breaks down. Renormalization through normal ordering with respect to the adiabatic vacuum state is required to extract finite results~\cite{Chung:2004nh, Kolb:2023ydq}. The procedure involves expressing the rotated operators $A_k$ and $A_{-k}^\dagger$ in terms of the original operators $\hat{a}_k$ and $\hat{a}_{-k}^\dagger$ using the inverse Bogoliubov transformation from Eq.~(\ref{eq:rotbog1}). The resulting normal-ordered expressions are as follows:
\begin{align}
    &: a_k a_{k'} : \; = \; a_k a_{k'} + \alpha_k^* \beta_k^* \delta^{(3)}(\mathbf{k} + \mathbf{k}') \, , \\
    & : a_k a_{-k'}^{\dagger} : \; = \;  a_{-k'}^{\dagger} a_k - |\beta_k|^2 \delta^{(3)}(\mathbf{k} + \mathbf{k}') \, , \\
    & : a_{-k'}^{\dagger} a_k : \; = \; a_{-k'}^{\dagger} a_k - |\beta_k|^2 \delta^{(3)}(\mathbf{k} + \mathbf{k}') \, , \\
    & : a_{-k'}^{\dagger} a_k^{\dagger} : \; = \; a_{-k'}^{\dagger} a_k^{\dagger} + \alpha_k \beta_k \delta^{(3)}(\mathbf{k} + \mathbf{k}') \, .
\end{align}

After normal ordering with respect to the adiabatic vacuum state and taking the vacuum expectation value, the renormalized energy density becomes:
\begin{equation}
\begin{aligned}
    \rho_{\chi}(\eta) = \langle : T_{00}^{(\chi)}(\eta, \mathbf{x}): \rangle = \frac{1}{2a^4} \int \frac{d^3k}{(2\pi)^3} \bigg[
        \mathcal{C}_1(k, -k) \alpha_k^* \beta_k^* 
        - \mathcal{C}_2(k, -k) |\beta_k|^2 - \mathcal{C}_3(k, -k) |\beta_k|^2 
        + \mathcal{C}_4(k, -k) \alpha_k \beta_k
    \bigg] \, ,
\end{aligned}
\end{equation}
where the $\mathcal{C}_i(k)$ coefficients arise from the normal ordering transformation and are linear combinations of the $\mathcal{B}_i(k, k')$ terms, given by
\begin{equation}
\begin{aligned}
    \label{appeq:cexpressions}
    \mathcal{C}_1(k, k') &= \mathcal{B}_1(k, k') \alpha_k \alpha_{k'} + \mathcal{B}_2(k, k') \alpha_k \beta_{k'} + \mathcal{B}_3(k, k') \alpha_{k'} \beta_k + \mathcal{B}_4(k, k') \beta_k \beta_{k'} \, , \\
    \mathcal{C}_2(k, k') &= \mathcal{B}_1(k, k') \alpha_k \beta^*_{k'} + \mathcal{B}_2(k, k') \alpha_k \alpha^*_{k'} + \mathcal{B}_3(k, k') \beta_k \beta^*_{k'} + \mathcal{B}_4(k, k') \alpha^*_{k'} \beta_k \, ,\\
    \mathcal{C}_3(k, k') &= \mathcal{B}_1(k, k') \alpha_{k'} \beta^*_{k} + \mathcal{B}_2(k, k') \beta_{k'} \beta^*_{k} + \mathcal{B}_3(k, k') \alpha^*_{k} \alpha_{k'} + \mathcal{B}_4(k, k') \alpha^*_{k} \beta_{k'} \, ,\\
    \mathcal{C}_4(k, k') &= \mathcal{B}_1(k, k') \beta^*_{k} \beta^*_{k'} + \mathcal{B}_2(k, k') \alpha^*_{k'} \beta^*_{k} + \mathcal{B}_3(k, k') \alpha^*_{k} \beta^*_{k'} + \mathcal{B}_4(k, k') \alpha^*_{k} \alpha^*_{k'} \, .
\end{aligned}
\end{equation}

Using the explicit forms of the Bogoliubov coefficients:
\begin{equation}
    \alpha_k \; = \; \frac{\omega_k^* X_k + i X_k'}{2 \omega_k f_k } \, , \qquad \beta_k \; = \; \frac{\omega_k X_k -i X_k'}{2 \omega_k f_k^* } \, , 
\end{equation}
the normal-ordered energy density can be expressed directly in terms of the mode functions:
\begin{equation}
\begin{aligned}
    \rho_{\chi} (\eta) \; = \; \frac{1}{2a^2} \int \frac{d^3 k}{(2\pi)^3} \bigg[ |\omega_k X_k - i X_k'|^2 + \left(\mathcal{H}^2 - \frac{a^2 R}{6} \right) |X_k|^2 - \mathcal{H} (X_k X_k'^* + X_k' X_k^* )- \frac{1}{2\omega_k} \left(\mathcal{H}^2 - \frac{a^2 R}{6} \right)  \bigg] \, .
\end{aligned}
\end{equation}

\section{Isocurvature Computation}
\label{app:B}
This section provides the detailed derivation of the isocurvature power spectrum used in the main text. The isocurvature power spectrum is computed from the two-point correlation function of energy density fluctuations:
\begin{equation}
    \langle \delta \rho_{\chi}(\eta, \mathbf{x}) \delta \rho_{\chi}(\eta, \mathbf{y}) \rangle = \langle :\rho_{\chi}(\eta, \mathbf{x}): :\rho_{\chi}(\eta, \mathbf{y}): \rangle - \bar{\rho}_{\chi}^2(\eta) \,.
\end{equation}
The normal-ordered product of energy density operators can be expressed as:
\begin{equation}
\begin{aligned}
    &:\rho_{\chi}(\eta, \mathbf{x}): :\rho_{\chi}(\eta, \mathbf{y}):\; = \;  \; = \; :T_{00}^{(\chi)}(\eta, \mathbf{x}): :T_{00}^{(\chi)}(\eta, \mathbf{y}): \; = \; \\
    &\frac{1}{4a^8} \int \frac{d^3 k \, d^3 k'}{(2\pi)^3} 
    \bigg[ \mathcal{C}_1(k, k') :a_k a_{k'}: + \, \mathcal{C}_2(k, k') :a_k a_{-k'}^\dagger: + \, \mathcal{C}_3(k, k') :a_{-k}^\dagger a_{k'}: + \, \mathcal{C}_4(k, k') :a_{-k}^\dagger a_{-k'}^\dagger: \bigg] \\
    & \times \int \frac{d^3 q \, d^3 q'}{(2\pi)^3} 
    \bigg[ \mathcal{C}_1(q, q') :a_q a_{q'}: + \, \mathcal{C}_2(q, q') :a_q a_{-q'}^\dagger: + \, \mathcal{C}_3(q, q') :a_{-q}^\dagger a_{q'}: + \, \mathcal{C}_4(q, q') :a_{-q}^\dagger a_{-q'}^\dagger: \bigg] \, ,
\end{aligned}
\end{equation}
where the coefficients $C_i(k, k')$ are given by Eq.~(\ref{appeq:cexpressions}). When taking the vacuum expectation value, most terms vanish due to the normal ordering, except for specific combinations that create particle pairs. The surviving terms correspond to contractions between creation and annihilation operators:
\begin{equation}
    \langle \mathcal{C}_1(k, k')\mathcal{C}_4(q, q') a_k a_{k'} a_{-q}^{\dagger} a_{-q'}^{\dagger} \rangle \; = \; \mathcal{C}_1(k, k')\mathcal{C}_4(q, q') \left[\delta^{(3)}(\mathbf{k}+\mathbf{q})\delta^{(3)}(\mathbf{k}'+\mathbf{q}') + \delta^{(3)}(\mathbf{k}+\mathbf{q}')\delta^{(3)}(\mathbf{k}'+\mathbf{q}) \right] \, .
\end{equation}
Performing the integrals with respect to $q$ and $q'$, we will be left with the contribution that is proportional to $\mathcal{C}_1(k, k') \mathcal{C}_4(-k, -k') + \mathcal{C}_1(k, k') \mathcal{C}_4(-k', -k)$. Therefore, we find that the full isocurvature expression is given by\footnote{Here, we relabeled $k \rightarrow k'$ and $k' \rightarrow k''$.}
\begin{equation}
\begin{aligned}
    &\Delta_{\mathcal S}^2 (\eta, k) \; = \; \frac{1}{\bar{\rho}_{\chi}^2(\eta)} \frac{1}{2a^8} \frac{k^3}{2\pi^2} \int d^3 \mathbf{r} \int \frac{d^3 k'}{(2\pi)^3}  \int \frac{d^3 k''}{(2\pi)^3} \times e^{i (\mathbf{k'} + \mathbf{k''} - \mathbf{k}) \cdot \mathbf{r}}\\
    &\times \Bigg( X_{k'} X_{k''} \left( a^2 m_{\text{eff}}^2 - k' k'' + \left(\mathcal{H}^2 -\frac{a^2 R}{6}\right) \right) + X_{k'}' X_{k''}' - \mathcal{H} \left( X_{k''} X_{k'}' + X_{k'} X_{k''}' \right) \Bigg) \\
    & \times \Bigg( X_{k'}^* X_{k''}^* \left( a^2 m_{\text{eff}}^2 - k' k'' + \left(\mathcal{H}^2 -\frac{a^2 R}{6}\right) \right) + X_{k'}^{*'} X_{k''}^{*'} - \mathcal{H}\left( X_{k''}^* X_{k'}^{*'} + X_{k'}^* X_{k''}^{*'} \right) \Bigg) \, .
\end{aligned}
\end{equation}
After performing integrals and relabeling $k' \rightarrow p$, we find:
\begin{equation}
\begin{aligned}
    \label{eq:isofullexpr}
    &\Delta_{\mathcal S}^2 (\eta, k) \; = \; \frac{1}{\bar{\rho}_{\chi}^2(\eta)} \frac{1}{2a^8} \frac{k^3}{2\pi^2} \int \frac{d^3 p}{(2\pi)^3} \\
    &\times \Bigg( X_{p} X_{q} \left( a^2 m_{\text{eff}}^2 - p q + \left(\mathcal{H}^2 -\frac{a^2 R}{6} \right) \right) + X_{p}' X_{q}' - \mathcal{H} \left( X_{p}' X_{q} + X_{p} X_{q}' \right) \Bigg) \\
    & \times \Bigg( X_{p}^* X_{q}^* \left( a^2 m_{\text{eff}}^2 - p q + \left(\mathcal{H}^2-\frac{a^2 R}{6}\right) \right) + X_{p}'^{*} X_{q}'^{*} - \mathcal{H} \left( X_{p}'^{*} X_{q}^*  + X_{p}^* X_{q}'^{*} \right) \Bigg) \, ,
\end{aligned}
\end{equation}
where $q = |\mathbf{p - k}|$. In the late-time asymptotic limit where the spacetime approaches Minkowski ($\mathcal{H} \rightarrow 0$, $R \rightarrow 0$) and for modes with $pq \ll a^2 m_{\text{eff}}^2$, the isocurvature power spectrum simplifies to:
\begin{equation}
\begin{aligned}
    &\Delta_{\mathcal{S}}^2 (\eta, k) \; = \; \frac{1}{\bar{\rho}_{\chi}^2(\eta)} \frac{k^3}{a^8 (2\pi)^5 } \int d^3 p \bigg[|X_p'|^2 |X_q'|^2 + a^4 m_{\chi, \rm eff}^4 |X_p|^2 |X_q|^2 + a^2 m_{\chi, \rm eff}^2 \bigg[ (X_p X_p'^*)(X_q X_q'^*) + \text{h.c.} \bigg] \, ,
\end{aligned}
\end{equation}    
which matches previous results~\cite{Garcia:2023awt, Ling:2021zlj}. This asymptotic expression provides a good approximation for estimating the isocurvature power spectrum at CMB scales. However, for a complete analysis across all scales, particularly at small scales relevant for gravitational wave signatures, we employ the full expression~(\ref{eq:isofullexpr}) and evolve it numerically throughout the cosmological history.

\begin{figure*}[t!]
        \centering
        \includegraphics[width=\linewidth]{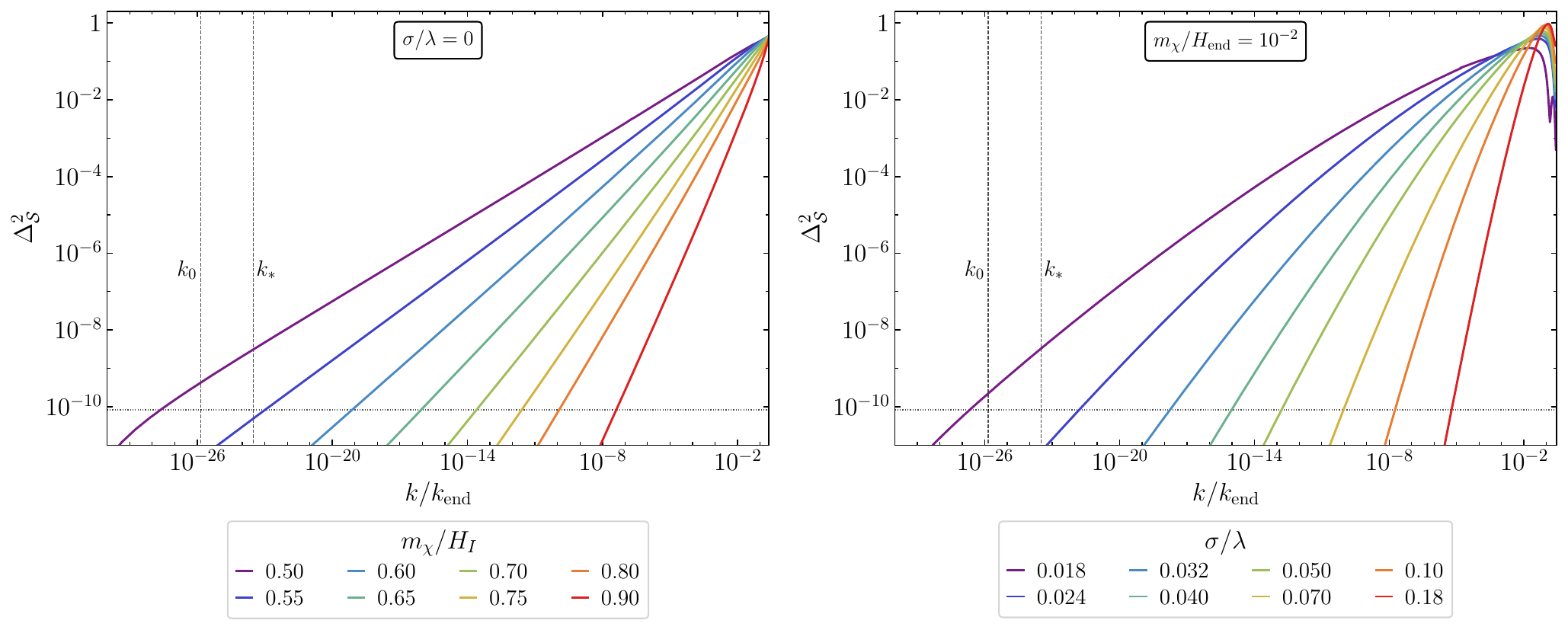}
        \caption{Isocurvature power spectrum for the spectator field $\chi$ for pure minimal gravitational production (left), and for a fixed mass with varying inflaton coupling (right). Each mass parameter (left panel) and coupling strength (right panel) is represented by a different color. Here $N_{\rm tot}=76.5$ $e$-folds of inflation. The vertical dashed lines show the positions of the present horizon scale ($k_0 = 1.4 \times 10^{-26} k_{\rm end}$) and the \textit{Planck} pivot scale ($k_* = 3.2\times 10^{-24}k_{\rm end}$). The horizontal dotted line represents the current upper bound on isocurvature perturbations from \textit{Planck}, $\Delta_{\mathcal{S}}^2 \simeq 8.3 \times 10^{-11}$ at the \textit{Planck} pivot scale.}
        \label{fig:iso1}
\end{figure*}

Figure~\ref{fig:iso1} shows the numerical isocurvature power spectrum for both purely gravitational production ($\sigma = 0$) and scenarios with non-zero inflaton-spectator coupling ($\sigma \neq 0$).

\section{Gravitational Wave Spectrum}
\label{app:gwspectrum}
This section details how scalar perturbations from spectator fields source secondary gravitational waves at second order in perturbation theory~\cite{Ananda:2006af, Baumann:2007zm, Domenech:2021ztg}. We review the general formalism and examine how the evolving isocurvature power spectrum influences the scalar potential $\Phi$, which subsequently sources tensor perturbations $h_{ij}$.

Working in the Newtonian gauge with the perturbed line element from Eq.~(\ref{eq:lineelement}), the equation of motion for secondary gravitational waves sourced by scalar perturbations is~\cite{Domenech:2021ztg}:
\begin{equation}
    \begin{aligned}
    \label{eq:hijincosmictime}
    &\ddot{h}_{ij} + 3H\dot{h}_{ij} - \frac{\nabla^2 h_{ij}}{a^2} \; = \; a^{-2} P^{ab}_{~~ij} \bigg\{4 \partial_a \Phi \partial_b \Phi + \frac{2}{M_P^2} \partial_a \delta \chi \partial_b \delta \chi \bigg\} \, ,
    \end{aligned}
\end{equation}
where $P^{ab}_{~~ij}$ is the transverse-traceless projection operator and $\delta \chi$ represents spectator field fluctuations. The inflaton perturbation $\delta \phi$ is negligible compared to the spectator field contribution since $\rho_\phi \gg \rho_\chi$ during the relevant epoch.

We work in the Newtonian gauge for scalar perturbations, with the line element defined in Eq.~(\ref{eq:lineelement}). The equation of motion for secondary GWs sourced by scalar perturbations is given by~\cite{Domenech:2021ztg}:
\begin{equation}
    \begin{aligned}
    \label{eq:hijincosmictime}
    &\ddot{h}_{ij} + 3H\dot{h}_{ij} - \frac{\nabla^2 h_{ij}}{a^2} \; = \; a^{-2} P^{ab}_{~~ij} \bigg\{4 \partial_a \Phi \partial_b \Phi + \frac{2}{M_P^2} \partial_a \delta \chi \partial_b \delta \chi \bigg\} \, ,
    \end{aligned}
\end{equation}
where $P^{ab}_{~~ij}$ is the projection operator, and $\delta \chi$ represents the dark matter fluctuations. Here, we neglected the $\delta \phi$ contribution since it is negligible compared to $\chi$ at the background level. Converting to conformal time using $dt = a d\eta$, the equation becomes:
\begin{equation}
    {h}_{ij}'' + 2\mathcal{H} h'_{ij} - \nabla^2 h_{ij} \; = \; P^{ab}_{~~ij} \left\{4 \partial_a \Phi \partial_b \Phi + \frac{2}{M_P^2} \partial_a \chi \partial_b \chi \right\} \, ,
\end{equation}
where primes denote derivatives with respect to conformal time $\eta$. The gravitational wave source consists of two distinct contributions:
\\

\textbf{Scalar potential term ($4 \partial_a \Phi \partial_b \Phi$)}. This arises from the nonlinear evolution of curvature perturbations enhanced by isocurvature fluctuations. The scalar potential $\Phi$ is related to density perturbations through the Poisson equation, creating a direct link between isocurvature power spectrum and gravitational wave production.
\\

\textbf{Direct field fluctuations ($\frac{2}{M_P^2} \partial_a \chi \partial_b \chi$)}. This term represents direct coupling of spectator field gradients to the gravitational wave equation. However, this contribution typically generates very high-frequency signals that fall outside the sensitivity range of current gravitational wave detectors and would require specialized resonant cavity experiments for detection.

For the present analysis, we focus exclusively on gravitational waves induced by the scalar potential term, which dominates in the frequency ranges accessible to pulsar timing arrays, space-based interferometers, and ground-based detectors. The computation of the resulting gravitational wave spectrum is detailed in the main text. To compute the tensor perturbations $h_{ij}$ and resulting gravitational wave spectra, we transform to Fourier space and employ Green's function methods. The energy density fluctuations are decomposed as:
\begin{equation}
    \delta \rho_{\chi}(\eta, \mathbf{x}) \; = \; \int \frac{d^3 k}{(2\pi)^{3/2}} e^{-i \mathbf{k} \cdot \mathbf{x}} \delta \rho_{\chi}(\eta, \mathbf{k}) \, .
\end{equation}
Using the gauge-invariant Poisson equation $\nabla^2 \Phi = \delta \rho_{\chi}/(2M_P^2)$ and projecting onto the transverse-traceless polarization tensors $\varepsilon^{\lambda}_{ij}(\mathbf{k})$, the gravitational wave equation becomes:
\begin{equation}
    \label{eq:hijgensol}
    h_{\mathbf{k}, \lambda}''(\eta) + 2\mathcal{H}  h_{\mathbf{k}, \lambda}'(\eta) + k^2 h_{\mathbf{k}, \lambda}(\eta) \; = \; S_{\lambda}(\eta, \mathbf{k}) \, ,
\end{equation}
where the source term $S_\lambda(\eta, \mathbf{k})$ is given by
\begin{equation}
\begin{aligned}
\label{eq:sourceterm}
    S_{\lambda}(\eta, \mathbf{k}) = -\int \frac{d^3 p}{(2\pi)^3} \frac{a^4}{M_P^4} \frac{\varepsilon_\lambda^{ab} p_a q_b}{p^2 q^2} \delta \rho_\chi(\eta, \mathbf{p}) \delta \rho_\chi(\eta, \mathbf{q}) \, .
\end{aligned}
\end{equation}
To solve the inhomogeneous differential equation~(\ref{eq:hijgensol}), we construct the retarded Green's function from two linearly independent solutions $h_1(\eta, k)$ and $h_2(\eta, k)$ of the homogeneous equation:
\begin{equation}
\mathcal{G}_{\mathbf k}(\eta, \tilde{\eta}) = \frac{1}{W(h_1, h_2, \tilde{\eta})} 
\left[ h_1(\eta) h_2(\tilde{\eta}) - h_1(\tilde{\eta}) h_2(\eta) \right],
\end{equation}
where the Green's function satisfies:
\begin{equation}
\mathcal{G}_{\mathbf{k}}''(\eta, \tilde{\eta}) + 2 \mathcal{H} \mathcal{G}_{\mathbf{k}}'(\eta, \tilde{\eta}) + k^2 \mathcal{G}_{\mathbf{k}}(\eta, \tilde{\eta}) = \delta(\eta - \tilde{\eta}) \, .
\end{equation}
The Wronskian determinant ensures the proper normalization:
\begin{equation}
W(h_1, h_2, \tilde{\eta}) \; = \;  h_1'(\tilde{\eta}) h_2(\tilde{\eta}) - h_1(\tilde{\eta}) h_2'(\tilde{\eta}) \, .
\end{equation}
With vanishing initial conditions $h_\lambda(\eta_i, \mathbf{k}) = h_\lambda'(\eta_i, \mathbf{k}) = 0$ at some early time $\eta_i$ before significant source contribution, the particular solution is obtained through convolution:
\begin{equation}
h_\lambda(\eta, \mathbf{k}) \; = \;  \int_{\eta_i}^\eta d\tilde{\eta} \, \mathcal{G}_{\mathbf k}(\eta, \tilde{\eta}) S_\lambda(\tilde{\eta}, \mathbf{k}) \, .
\end{equation}
This formulation provides the foundation for our numerical computation of the gravitational wave amplitude $h_\lambda(\eta, \mathbf{k})$, from which we derive the energy density spectrum $\Omega_{\text{GW}}(k)$ presented in the main text.

Next, we follow the methodology developed in Refs.~\cite{Garcia-Saenz:2022tzu, Adshead:2021hnm, Unal:2018yaa, Atal:2021jyo}. We assume statistical homogeneity and isotropy of the spectator field density perturbations $\delta \rho_{\chi}$ that appear in the gravitational wave source term~(\ref{eq:sourceterm}). With vanishing mean $\langle \delta \rho_{\chi} \rangle = 0$, the two-point correlation function of induced gravitational waves can be expressed as:
\begin{equation}
\begin{aligned}
\label{eq:4pointcorr1}
\langle h_\lambda(\eta, \mathbf{k}_1) h_\lambda(\eta, \mathbf{k}_2) \rangle \; \sim \; \int \frac{d^3 q_1}{(2\pi)^{3/2}} \frac{d^3 q_2}{(2\pi)^{3/2}}  \, \langle \delta \rho_{\chi}(\mathbf{q}_1) \delta \rho_{\chi}(\mathbf{k}_1 - \mathbf{q}_1) \delta \rho_{\chi}(\mathbf{q}_2) \delta \rho_{\chi}(\mathbf{k}_2 - \mathbf{q}_2) \rangle \, .
\end{aligned}
\end{equation}
This expression involves a four-point correlation function of density perturbations, which contains both Gaussian and non-Gaussian contributions. The four-point correlation function can be decomposed into disconnected and connected parts:
\begin{equation}
\begin{aligned}
    \langle  \delta \rho_{\chi}(\mathbf{k}_1)  \delta \rho_{\chi}(\mathbf{k}_2)  \delta \rho_{\chi}(\mathbf{k}_3)  \delta \rho_{\chi}(\mathbf{k}_4) \rangle \; = \; \langle  \delta \rho_{\chi}(\mathbf{k}_1)  \delta \rho_{\chi}(\mathbf{k}_2)  \delta \rho_{\chi}(\mathbf{k}_3)  \delta \rho_{\chi}(\mathbf{k}_4) \rangle_d 
    + \langle  \delta \rho_{\chi}(\mathbf{k}_1)  \delta \rho_{\chi}(\mathbf{k}_2)  \delta \rho_{\chi}(\mathbf{k}_3)  \delta \rho_{\chi}(\mathbf{k}_4) \rangle_c \, ,
\end{aligned}   
\end{equation}
where the disconnected part represents the Gaussian contribution and the connected part captures non-Gaussian effects. These contributions are given by:
\begin{equation}
    \begin{aligned}
        \langle \delta \rho_{\chi}(\mathbf{k}_1) \delta \rho_{\chi}(\mathbf{k}_2) \delta \rho_{\chi}(\mathbf{k}_3) \delta \rho_{\chi}(\mathbf{k}_4) \rangle_c \; = \; \delta^{(3)}(\mathbf{k}_1 + \mathbf{k}_2 + \mathbf{k}_3 + \mathbf{k}_4) 
    \mathcal{T}(\mathbf{k}_1, \mathbf{k}_2, \mathbf{k}_3, \mathbf{k}_4) \,,
    \end{aligned}
\end{equation}
and
\begin{equation}
    \begin{aligned}
        &\langle \delta \rho_{\chi}(\mathbf{k}_1) \delta \rho_{\chi}(\mathbf{k}_2) \delta \rho_{\chi}(\mathbf{k}_3) \delta \rho_{\chi}(\mathbf{k}_4) \rangle_d \; = \;  \langle \delta \rho_{\chi}(\mathbf{k}_1) \delta \rho_{\chi}(\mathbf{k}_2) \rangle \langle \delta \rho_{\chi}(\mathbf{k}_3) \delta \rho_{\chi}(\mathbf{k}_4) \rangle \\
        &+ \langle \delta \rho_{\chi}(\mathbf{k}_2) \delta \rho_{\chi}(\mathbf{k}_3) \rangle \langle \delta \rho_{\chi}(\mathbf{k}_4) \delta \rho_{\chi}(\mathbf{k}_1) \rangle 
        + \langle \delta \rho_{\chi}(\mathbf{k}_1) \delta \rho_{\chi}(\mathbf{k}_3) \rangle \langle \delta \rho_{\chi}(\mathbf{k}_2) \delta \rho_{\chi}(\mathbf{k}_4) \rangle \, .
    \end{aligned}
\end{equation}
Here, $\mathcal{T}(\mathbf{k}_1, \mathbf{k}_2, \mathbf{k}_3, \mathbf{k}_4)$ is the connected trispectrum, and the two-point function is defined as
\begin{equation}
\langle \delta \rho_{\chi}(\mathbf{k}_1) \delta \rho_{\chi}(\mathbf{k}_2) \rangle \; =  \;  \delta^{(3)}(\mathbf{k}_1 + \mathbf{k}_2) \bar{\rho}_{\chi}^2 \mathcal{P}_{\mathcal{S}}(\mathbf{k}_1) \, .
\end{equation} 

For spectator field models with weak primordial non-Gaussianities, we treat the connected contribution as a small perturbative correction to the dominant Gaussian statistics. The gravitational wave power spectrum accordingly splits as:
\begin{equation}
    \mathcal{P}_\lambda(k) \; = \; \mathcal{P}_\lambda(k)|_d + \mathcal{P}_\lambda(k)|_c \, ,
\end{equation}
where the Gaussian piece arises from products of two-point functions (disconnected diagrams) and the non-Gaussian piece encodes contributions from the primordial trispectrum (connected diagrams). Since the connected contribution is typically subdominant for the parameter ranges of interest, we focus on computing the Gaussian contribution. A comprehensive analysis including non-Gaussian effects is provided in Ref.~\cite{Garcia-Saenz:2022tzu}.

The Gaussian approximation requires evaluating the disconnected part of the four-point correlation function appearing in Eq.~(\ref{eq:4pointcorr1}):
\begin{equation}
    \begin{aligned}
        & \langle \delta \rho_{\chi}(\mathbf{q}_1) \delta \rho_{\chi}(\mathbf{k}_1 - \mathbf{q}_1) \delta \rho_{\chi}(\mathbf{q}_2) \delta \rho_{\chi}(\mathbf{k}_2 - \mathbf{q}_2) \rangle_d \; = \; \langle \delta \rho_{\chi}(\mathbf{q}_1) \delta \rho_{\chi}(\mathbf{k}_1 - \mathbf{q}_1)  \rangle \langle \delta \rho_{\chi}(\mathbf{q}_2) \delta \rho_{\chi}(\mathbf{k}_2 - \mathbf{q}_2)  \rangle \\
        &+ \langle \delta \rho_{\chi}(\mathbf{k}_1 - \mathbf{q}_1)  \delta \rho_{\chi}(\mathbf{q}_2) \rangle \langle \delta \rho_{\chi}(\mathbf{k}_2 - \mathbf{q}_2) \delta \rho_{\chi}(\mathbf{q}_1) \rangle + \langle \delta \rho_{\chi}(\mathbf{q}_1) \delta \rho_{\chi}(\mathbf{q}_2) \rangle \langle \delta \rho_{\chi}(\mathbf{k}_1 - \mathbf{q}_1)  \delta \rho_{\chi}(\mathbf{k}_2 - \mathbf{q}_2) \rangle \, .
    \end{aligned}
\end{equation}
When substituting the explicit form of the two-point functions, certain terms generate contributions proportional to $\delta^{(3)}(\mathbf{k}_1)$ and $\delta^{(3)}(\mathbf{k}_2)$. These zero-momentum modes correspond to unphysical infrared divergences that must be regularized. Since we are interested in gravitational waves with finite momentum $k_1, k_2 \neq 0$, these terms can be consistently dropped. After this regularization procedure, the physical contribution to the disconnected four-point function becomes:
\begin{equation}
    \label{eq:4pointsplit}
    \begin{aligned}
        \langle \delta \rho_{\chi}(\mathbf{q}_1) \delta \rho_{\chi}(\mathbf{k}_1 - \mathbf{q}_1) \delta \rho_{\chi}(\mathbf{q}_2) \delta \rho_{\chi}(\mathbf{k}_2 - \mathbf{q}_2) \rangle_d \; = \; \delta^{(3)}(\mathbf{k}_1 + \mathbf{k}_2) 
        \big[ \delta^{(3)}(\mathbf{q}_1 + \mathbf{q}_2) + \delta^{(3)}(\mathbf{k}_1 + \mathbf{q}_2 - \mathbf{q}_1) \big] \bar{\rho}_{\chi}^4 \mathcal{P}_{S}(\mathbf{q}_1) \mathcal{P}_{S}(|\mathbf{k}_1 - \mathbf{q}_1|) \, .
    \end{aligned}
\end{equation}

The gravitational wave two-point correlation function is obtained by convolving the source correlations with the appropriate Green's functions:
\begin{equation}
\begin{aligned}
    \label{eq:gwspectrgenexp}
   \langle h_\lambda(\eta, \mathbf{k}_1) h_\lambda(\eta, \mathbf{k}_2) \rangle \; = \; \int_0^{\eta} d\eta_1 \int_0^{\eta} d\eta_2 \, \mathcal{G}_{\mathbf k}(\eta, \eta_1) \mathcal{G}_{\mathbf k}(\eta, \eta_2) \langle S_{ \lambda}(\eta_1, \mathbf{k}_1) S_{\lambda}(\eta_2, \mathbf{k}_2) \rangle \, ,
\end{aligned}
\end{equation}
where the source two-point function is given by:
\begin{equation}
\begin{aligned}
    \label{eq:s2point}
    \langle S_{ \lambda}(\eta_1, \mathbf{k}_1) S_{\lambda}(\eta_2, \mathbf{k}_2) \rangle & =  \frac{a(\eta_1)^4}{M_P^4} \frac{a(\eta_2)^4}{M_P^4} \int \frac{d^3q_1}{(2\pi)^{3/2}} \frac{d^3q_2}{(2\pi)^{3/2}} \frac{Q_{\lambda}(\mathbf{k}_1, \mathbf{q}_1) Q_{\lambda}(\mathbf{k}_2, \mathbf{q}_2)}{q_1^2 |\mathbf{k}_1 - \mathbf{q}_1|^2 q_2^2 |\mathbf{k}_2-\mathbf{q}_2|^2} \\
    &\times \mathcal{T}_{\mathcal{S}}(\eta_1) \mathcal{T}_{\mathcal{S}}(\eta_2) \langle \delta \rho_{\chi}(\mathbf{q}_1) \delta \rho_{\chi}(|\mathbf{k}_1-\mathbf{q}_1|) \delta \rho_{\chi}(\mathbf{q}_2) \delta \rho_{\chi}(|\mathbf{k}_2-\mathbf{q}_2|) \rangle \, .
\end{aligned}    
\end{equation}
Here, we used the property that we found from the numerical analysis:
\begin{equation}
\label{eq:TS}
\Delta_{\mathcal{S}}^2(\eta,k) \;=\; \mathcal{T}_{\mathcal{S}}(\eta)\Delta_{\mathcal{S}}^2(k)\,,\quad (k<aH) \, ,
\end{equation}
where $\mathcal{T}_{\mathcal{S}}(\eta)$ represents a time-dependent transfer function. The gravitational wave polarization structure is encoded in the projection factor:
\begin{equation}
    Q_{\lambda}(\mathbf{k}, \mathbf{q}) \; \equiv \; \varepsilon_{ij}^{\lambda}(\mathbf{k}) q_i q_j \; = \; -\varepsilon_{\lambda}^{ij}(\mathbf{k})(\mathbf{k}-\mathbf{q})_i q_j \, ,
\end{equation}
where $\varepsilon_{ij}^{\lambda}(\mathbf{k})$ are the transverse-traceless polarization tensors satisfying $\varepsilon_\lambda^{ij}(\mathbf{k}) k_i = 0$.
For computational convenience, we choose coordinates with $\hat{\mathbf{k}} = \hat{\mathbf{z}}$ and parameterize $\mathbf{q} = q(\sin \theta \cos \phi, \sin \theta \sin \phi, \cos \theta)$. The projection factors for the two polarization states become:
\begin{equation}
\begin{aligned}
\label{eq:qexpressions}
&Q_+(\mathbf{k}, \mathbf{q}) \; = \; \frac{q^2}{\sqrt{2}} \sin^2 \theta \cos(2\phi) \, , \\
&Q_\times(\mathbf{k}, \mathbf{q}) \; = \; \frac{q^2}{\sqrt{2}} \sin^2 \theta \sin(2\phi) \, .
\end{aligned}
\end{equation}
The projection factors possess useful symmetry properties that simplify the momentum integrals. Because $\varepsilon_\lambda(\mathbf{k})$ is orthogonal to $\mathbf{k}$, we find:
\begin{equation}
Q_\lambda(\mathbf{k}, \mathbf{q}) = Q_\lambda(\mathbf{k}, \mathbf{q} + c\mathbf{k}) \, ,
\end{equation}
for any constant $c$. $Q_\lambda(\mathbf{k}, \mathbf{q})$ is symmetric under 
$\mathbf{k} \to -\mathbf{k}$ and $\mathbf{q} \to -\mathbf{q}$:
\begin{equation}
Q_\lambda(\mathbf{k}, \mathbf{q}) = Q_\lambda(-\mathbf{k}, \mathbf{q}) = Q_\lambda(\mathbf{k}, -\mathbf{q}) = Q_\lambda(-\mathbf{k}, -\mathbf{q}) \, .
\end{equation}

The GW power spectrum can be expressed as:
\begin{equation}
\label{eq:gwpowerspectrum2}
\langle h_\lambda(\eta, \mathbf{k}) h_{\lambda'}(\eta, \mathbf{k}') \rangle \; = \; \delta_{\lambda \lambda'} \delta^{(3)}(\mathbf{k} + \mathbf{k}') \mathcal{P}_\lambda(\eta, k) \,, 
\end{equation}
where the dimensionless power spectrum is given by:
\begin{equation}
\label{eq:gwdimensionlessspectrum}
\Delta^2_\lambda(\eta, k) \; = \; \frac{k^3}{2\pi^2} \mathcal{P}_\lambda(\eta, k) \, .
\end{equation}
The gravitational wave energy density is conventionally expressed as a fraction of the critical density $\rho_c = 3H^2M_P^2$:
\begin{equation}
\Omega_{\text{GW}}(k) \; \equiv\;  \frac{1}{3M_P^2H^2} \frac{d \rho_{\text{GW}}}{d\ln k} \, ,
\end{equation}
where $\rho_{\text{GW}}$ is the gravitational wave energy density. The relationship between the dimensionless power spectrum and the energy density parameter is:
\begin{equation}
\label{eq:gwgeneral}
\Omega_{\text{GW}}(k) \; = \; \frac{1}{12} \left( \frac{k}{a(\eta)H(\eta)} \right)^2 \sum_\lambda \Delta_\lambda^2 \,.
\end{equation}
For comparison with observational constraints, we evaluate the spectrum at the present epoch. Using the present-day Hubble parameter $H_0 = 100 \, h \, \text{km s}^{-1} \text{Mpc}^{-1}$ and scale factor $a_0$, the observable gravitational wave energy density becomes:
\begin{equation}
\label{eq:gwpresentday}
\Omega_{\text{GW},0}(k) h^2 \; = \; \frac{1}{12} \left( \frac{k}{a_0 H_0} \right)^2  \Delta_h^2 \,, 
\end{equation}
where $\sum_\lambda \Delta_\lambda^2 = \Delta_h^2$ is the total GW power spectrum across both polarization states.

The complete dimensionless gravitational wave power spectrum can be expressed in factorized form as:
\begin{equation}
\label{eq:deltah2}
\begin{aligned}
&\Delta_h^2(k, N) \;=\; 2 \left[\frac{1}{8}\int^NdN'\, \mathcal{G}_{\mathbf k}(N,N')\, \mathcal{T}_{\mathcal{S}}(N')\left(\frac{a(N')H(N')}{k}\right)^2\left(\frac{\rho_{\chi}(N')}{H^2(N')M_P^2}\right)^2\right]^2g(k)
\, ,
\end{aligned}
\end{equation}
where 
\begin{equation}
\label{eq:fgen}
g(k) \;=\; k^2\int_0^{\infty} dp\, p\int_{|k-p|}^{k+p}dq\, q\,\frac{(k^4-2k^2(p^2+q^2)+(p^2-q^2)^2)^2}{p^7q^7}\,\Delta_{\mathcal{S}}^2(p)\Delta_{\mathcal{S}}^2(q)\,.
\end{equation}
The gravitational wave energy density spectrum becomes:  
\begin{equation}
    \label{eq:gwiso}
    \Omega_{{\rm GW}, \, \mathcal{S}} \; = \; \frac{1}{12} \left(\frac{k}{a H} \right)^2 \Delta_h^2(k, N) \; \equiv \; \mathcal{I}(k, N) g(k) \, .
\end{equation}
This factorization separates the time-dependent gravitational wave growth from the momentum-space structure determined by the isocurvature spectrum. The calculation is performed using the number of $e$-folds $N$ as the time variable, which provides computational advantages during the inflationary and post-inflationary epochs where the scale factor evolves exponentially. Correspondingly, $\mathcal{G}_k(N,N')$ represents the Green's function expressed in $e$-fold time rather than conformal time. The normalization of $g(k)$ includes appropriate factors of $k$ to ensure dimensional consistency, specifically, the $k^2$ prefactor in Eq.~(\ref{eq:fgen}) renders $g(k)$ dimensionless while preserving the correct scaling behavior of the gravitational wave spectrum.
\newpage

\end{document}